\documentclass[fleqn,usenatbib]{mnras}

\usepackage{newtxtext,newtxmath}

\usepackage[T1]{fontenc}
\usepackage{ae,aecompl}

\usepackage{graphicx}	
\usepackage{amsmath}	
\usepackage{amssymb}	
\usepackage{hyperref}
\usepackage{caption}
\usepackage{pdflscape}

\usepackage{siunitx}
\usepackage{xcolor}
\definecolor{alizarin}{rgb}{0.82, 0.1, 0.26}

\begin{document}
\title[Resolving quenching in the green valley]{What Drives Galaxy Quenching? Resolving Molecular Gas and Star Formation in the Green Valley}

\author[S Brownson et al.]{Simcha Brownson$^{1,2}$\thanks{E-mail: sbb33@cam.ac.uk}, Francesco Belfiore$^{3, 4}$, Roberto Maiolino$^{1,2}$, Lihwai Lin$^{5}$ \newauthor 
and Stefano Carniani$^{6}$
\\
$^{1}$Kavli Institute for Cosmology, University of Cambridge, Madingley Road, Cambridge CB3 0HA, UK\\
$^{2}$Cavendish Laboratory, University of Cambridge, 19 J. J. Thomson Ave., Cambridge CB3 0HE, UK\\
$^{3}$European Southern Observatory, Karl-Schwarzschild-Str. 2, Garching bei M{\"u}nchen, D-85748, Germany\\
$^{4}$INAF -- Osservatorio Astrofisico di Arcetri, Largo E. Fermi 5, I-50157, Firenze, Italy\\
$^{5}$Institute of Astronomy and Astrophysics, Academia Sinica, Taipei 10617, Taiwan\\
$^{6}$Scuola Normale Superiore, Piazza dei Cavalieri 7, I-56126 Pisa, Italy\\
}

\date{Accepted XXX. Received YYY; in original form ZZZ}

\pubyear{2018}


\label{firstpage}
\pagerange{\pageref{firstpage}--\pageref{lastpage}}
\maketitle

\begin{abstract}
We study quenching in seven green valley galaxies on kpc scales by resolving their molecular gas content using \textsuperscript{12}CO(1-0) observations obtained with NOEMA and ALMA, and their star-formation rate using spatially resolved optical spectroscopy from the MaNGA survey. We perform radial stacking of both datasets to increase the sensitivity to  molecular gas and star formation, thereby avoiding biases against strongly quenched regions. We find that both spatially resolved gas fraction ($\rm {f_{gas}}$) and star formation efficiency ($\rm {SFE}$) are responsible for quenching green valley galaxies at all radii: both quantities are suppressed with respect to typical  star-forming regions. $\rm {f_{gas}}$ and $\rm {SFE}$ have roughly equal influence in quenching the outer disc. We are, however, unable to identify the dominant mechanism in the strongly quenched central regions.  We find that $\rm {f_{gas}}$ is reduced by $\rm \sim 1~dex$ in the central regions, but the star formation rate is too low to be measured, leading to upper limits for the $\rm {SFE}$. Moving from the outer disc to central regions, the reduction in $\rm {f_{gas}}$ is driven by an increasing $\rm \Sigma_{\star}$ profile rather than a decreasing $\rm \Sigma_{H_{2}}$ profile. The reduced $\rm {f_{gas}}$  may therefore be caused by a decrease in the gas supply rather than molecular gas ejection mechanisms, such as winds driven by active galactic nuclei. We warn more generally that studies investigating $\rm {f_{gas}}$ may be deceiving in inferring the cause of quenching, particularly in the central (bulge-dominated) regions of galaxies.
\end{abstract}

\begin{keywords}
Galaxies: evolution, galaxies: star formation, galaxies: general
\end{keywords}




\section{Introduction}\label{Introduction}
Star forming and passive galaxies differ in key properties, such as colour, morphology, and star formation rate (SFR, \citealt{Strateva2001, Springel2004, Baldry2004, Renzini2015}). Galaxies in the green valley (GV) region of the colour-magnitude diagram have intermediate properties and  the majority of these are thought to be transitioning from being blue and star-forming to red and passive \citep{Wyder2007, Martin2007}, a process commonly referred to as quenching. 

The advent of large optical integral field unit (IFU) surveys is enabling spatially resolved studies of the physics governing galaxy quenching. For example, outside-in quenching models (e.g. ram-pressure stripping, e.g. \citealt{Kenney2004}) can be tested against inside-out models (e.g. feedback from active galactic nuclei [AGN], e.g. \citealt{Fabian2012}) by resolving the spatial distribution of star formation. One such study of spatially resolved star formation demonstrated that massive GV galaxies host central low-ionisation emission-line regions (cLIERs,  \citealt{Belfiore2017a}). These cLIER galaxies form stars in their outer discs, but their central emission is dominated by old stellar populations, indicating a lack of recent star formation. \citet{Belfiore2018} found that, although the quenching is most extreme in the central regions, star formation is suppressed at all radii: quenching does not simply occur inside-out.

Data from IFUs and sub-mm interferometers, with matched kpc-scale spatial resolution, can be combined to investigate the conversion of gas into stars, a process that is governed on local, spatially resolved scales \citep{Schinnerer2019}. The ALMA-MaNGA QUEnching and STar formation (ALMaQUEST) project is one of the first resolved studies to systematically investigate galaxies across the $\rm \Sigma_{\star} - \Sigma_{SFR}$ plane at $\rm z \sim 0$ (Lin et al., in prep). \citet{Lin2019} use a sample of star-forming ALMaQUEST galaxies to calibrate three resolved relationships: the spatially-resolved star formation main sequence (rSFMS, $\rm \Sigma_{\star} - \Sigma_{SFR}$) (e.g. \citealt{CanoDiaz2016}), the molecular gas main sequence (rMGMS, $\rm \Sigma_{\star} - \Sigma_{H_2}$), and the Schmidt-Kennicutt star formation law (rSK, $\rm \Sigma_{H_2} - \Sigma_{SFR}$) \citep{Kennicutt1998}. Offsets from these relationships can therefore be used to quantify quenching in the GV on kpc-scales.

In this letter we investigate quenching of star formation by comparing the distribution of molecular gas and star formation in a sample of seven massive GV galaxies. Five galaxies were selected to lie in the GV in  $NUV-r$ colours (4 < $NUV-r$ < 5), to have large central 4000~\AA~ breaks (indicative of the old central stellar populations found in bulges),  not to host a Seyfert AGN, and to have axis ratios larger than 0.5 to avoid inclination effects. The large selected central 4000~\AA~ breaks are representative of massive ($\rm M_{\star}>10^{10}M_{\odot}$) GV galaxies, lying within 1~$\sigma$ of the population mean. We also reanalyse two GV galaxies without AGN  from the \citet{Lin2017} ALMaQUEST pilot study. This work uses a larger sample size than the pilot study, and performs a radial stacking analysis to avoid biases due to non-detection of either SFR or molecular gas tracers.  We assume a \citet{Kroupa2001} IMF and  $\rm \Lambda$CDM cosmology  throughout, with $\rm H_{0}$ = 70  $\rm {km} \ \rm{ s}^{-1}$ $\rm Mpc{}^{-1}$, $\rm {\Omega}_{M}$ = 0.3 and $\rm {\Omega}_{\Lambda}$ = 0.7.

\section{Data}\label{Data}

\subsection{MaNGA integral field spectroscopy}
\label{MaNGA Observations}
Mapping Nearby Galaxies at Apache Point Observatory (MaNGA) is an IFU survey targeting 10,000 nearby galaxies (z$\sim$0.03, \citealt{Bundy2015, Yan2016}). Mounted on the SDSS 2.5~m telescope \citep{Gunn2006}, the IFU system simultaneously targets 17 galaxies, covering them out to at least 1.5~effective radii ($\rm R_{e}$). The fibres are fed into the BOSS spectrographs \citep{Smee2013}, which fully cover the wavelength range 3600-10000~\AA\ with spectral resolution R$\sim$2000. Reduced data cubes have 0.5~arcsec spaxels and a spatial resolution (full width at half maximum) of 2.5~arcsec  \citep{Yan2016b, Law2016}. The MaNGA data used in this work is taken from data release 15 \citep{Aguado2019}.

We analyse the data both spaxel-by-spaxel and in bins of deprojected radius, generated using the position angles and inclinations from the NASA-Sloan catalogue \citep{Blanton2011}, derived from SDSS photometry. The spaxel-by-spaxel analysis is used to obtain an initial view of the data (as shown in Fig. \ref{fig:ExampleResolvedMaps}) and to obtain the velocity field used to stack spectra in radial bins. We describe the stacking analysis in detail below since it forms the basis of our result. The spaxel-by-spaxel analysis follows roughly the same steps.

We first recenter and coadd the spectra of spaxels in bins of width 0.25 $\rm R_{e}$ using the H$\alpha$ velocity field from the data analysis pipeline (\textsc{dap}) v2.2.1 \citep{Westfall2019, Belfiore2019}. We construct a grid of  72 simple stellar population (SSP) templates spanning 12 ages (0.001 to 15 Gyr) and six metallicities ([Z/H] = -2.0 to 0.0) using  the  \textsc{pegase}-\textsc{hr} code \citep{LeBorgne2004} together with the Elodie v3.1 stellar library \citep{Prugniel2001, Prugniel2007}, and then use penalised pixel fitting (\textsc{ppxf}, \citealt{Cappellari2004, Cappellari2017}) to simultaneously fit the gas and stellar emission whilst assuming a \citet{Calzetti2001} attenuation curve. We refit the spectra after adding noise, producing a distribution of 1000 estimates for the emission line fluxes and the mass in each SSP template. 
We have checked that the H$\alpha$ fluxes obtained in this way are consistent with those obtained by summing the individual spaxel flux estimates from the \textsc{dap}. The H$\alpha$ flux is corrected for dust extinction using the theoretical case B Balmer ratio (H$\alpha$/H$\beta$ = 2.87) and the \citet{Calzetti2001} attenuation curve with $R_V$ = 4.05. $\rm \Sigma_{SFR}$  is derived from the extinction-corrected H$\alpha$ flux using the \citet{Kennicutt2012} calibrations for a \citet{Kroupa2001} IMF for spectra classified as star-forming in the [S\textsc{ii}]$\lambda$6717,31/H$\alpha$ ([S\textsc{ii}]/H$\alpha$) versus [O\textsc{iii}]$\lambda$5007/H$\beta$ ([O\textsc{iii}]/H$\beta$)  Baldwin-Phillips-Terlevich (BPT) diagram \citep{Baldwin1981, Veilleux1987}. We have checked that the radial and spaxel-by-spaxel BPT classifications are consistent; less than 10 per cent of the spaxels in LIER radial bins are star forming.

$\rm \Sigma_{\star}$ is estimated from the average reconstructed star formation history in each spaxel, defined as the mean mass over all MC runs in each age slice. We correct $\Sigma_{\star}$ for the mass fraction returned to the ISM. In regions which are BPT-classified as LIER we also use the SSP analysis to test for the presence of young stars. We define $\rm \Sigma_{SFR}$ in LIER regions as the average rate of star formation in the last 10~Myr, consistent with the star formation timescale probed by H$\alpha$ \citep{Kennicutt2012}.
We define a conservative sensitivity limit to young stars using the 10$\rm ^{th}$ percentile of spatially resolved $\rm {sSFR}$ for all annular fits with non-zero weights for young stars: log($\rm sSFR/yr^{-1})~ \sim~-12$. We choose an $\rm {sSFR}$ limit, rather than $\rm \Sigma_{SFR}$,  since the sensitivity to young stars is strongly affected by the total mass budget. The sensitivity limit is combined with $\rm \Sigma_{\star}$ to place constraining upper limits on $\rm \Sigma_{SFR}$ in annuli lacking evidence of recent star formation.

\subsection{CO(1-0) data}
\textsuperscript{12}CO(1-0) observations have been performed for a sample of five galaxies using the NOrthern Extended Millimeter Array (NOEMA). Each galaxy was observed for $\sim$5.5~hr total on-source time in two array configurations: C (observed June-July 2017, typically with 10 < precipitable water vapor (PWV) < 15mm) and D (observed April 2018, typically with 5 < PWV < 10mm). Data reduction and imaging are performed using the \textsc{gildas} software packages \textsc{clic} and \textsc{mapping}. The absolute flux calibration at $\sim$100~GHz is typically precise to better than $\sim$10~per~cent. Dirty cubes with channel widths of 10.7~km~s$^{-1}$ are produced using natural weighting and cleaned down to the 1$\rm \sigma$ noise level  using the H\"{o}gbom deconvolving algorithm \citep{Hogbom1974}. The beam sizes are generally well-matched to the the MaNGA point spread function (PSF), except for 8604-12701 whose beam size is slightly larger due to a pointing error during the April 2018 observations (see Table A1 in the online supplementary material). 8604-12701's wide (3.1~arcsec) $\rm 0.25~R_{e}$ radial bins ensure that the stacked NOEMA and MaNGA data probe similar spatial scales, so we do not attempt to match the resolution in this galaxy.

The \textsuperscript{12}CO(1-0) flux in each pixel is estimated by integrating across the set of adjacent channels that maximises the signal-to-noise ratio (SNR), where the noise level is given by the root-mean-square flux of channels more than $\sim$350~km~s$^{-1}$ offset from the emission line centroid. The flux in spaxels with SNR below five is set to the 5$\sigma$ detection limit. We show an example of the maps obtained in this way in Fig. \ref{fig:ExampleResolvedMaps}. 

We base the analysis in this paper on radially-stacked profiles. In particular, we use the H$\alpha$ velocity field from the \textsc{dap} to coadd the NOEMA spectra of spaxels within annular bins of width 0.25~$\rm R_{e}$, and we measure the line flux by integrating across the channels above the 1$\rm \sigma$ noise level. Recentering the \textsuperscript{12}CO(1-0) emission line ensures that the coadded spectrum has a single peak rather than a double-horned profile and increases the SNR of the emission (See Fig. C1 in the online supplementary material). A standard Milky Way CO-to-$\rm H_2$ conversion factor ($\rm {\alpha}_{CO}$) of 4.3~M$_{\odot}$~(K~km~s$^{-1}$~pc$^{2}$)$^{-1}$ is used to calculate the H$_{2}$ mass surface density ($\rm {\Sigma}_{{H}_{2}}$) \citep{Bolatto2013}. 

For the two galaxies observed by ALMA and presented in \cite{Lin2017} (see bottom two rows of Table A1 in the online supplementary material), we follow the same analysis procedure as for our new NOEMA targets.

\section{Results}
\label{Results}

Quenching is the suppression of star formation (often quantified by $\rm {sSFR} = \Sigma_{SFR} / \Sigma_{{\star}} $, with `quenched' regions having $\rm {sSFR}~\sim 10^{-12}~{yr}^{-1}$, consistent with the population of passive galaxies)  and can occur because of a reduced molecular gas content, often quantified in terms of gas fraction $\rm {f_{gas}} = \Sigma_{H_{2}} / \Sigma_{{\star}} $, and/or star formation efficiency $\rm {SFE} = \Sigma_{SFR} /  \Sigma_{H_{2}}$, where
\begin{equation}
\label{EQN:sSFR,fgas,SFE}
    \rm
    log({sSFR}) = log({SFE}) + log({f_{gas}}).
\end{equation}
 In this framework, a reduction in $\rm {SFE}$ probes quenching through inefficient conversion of gas into stars while low $\rm {f_{gas}}$ signifies quenching through a depleted gas reservoir. We note that $\rm {sSFR},~{SFE},~and~{f_{gas}}$ denote spatially resolved quantities unless otherwise stated.  

\begin{figure}
\center
    \includegraphics[width=0.5\textwidth]{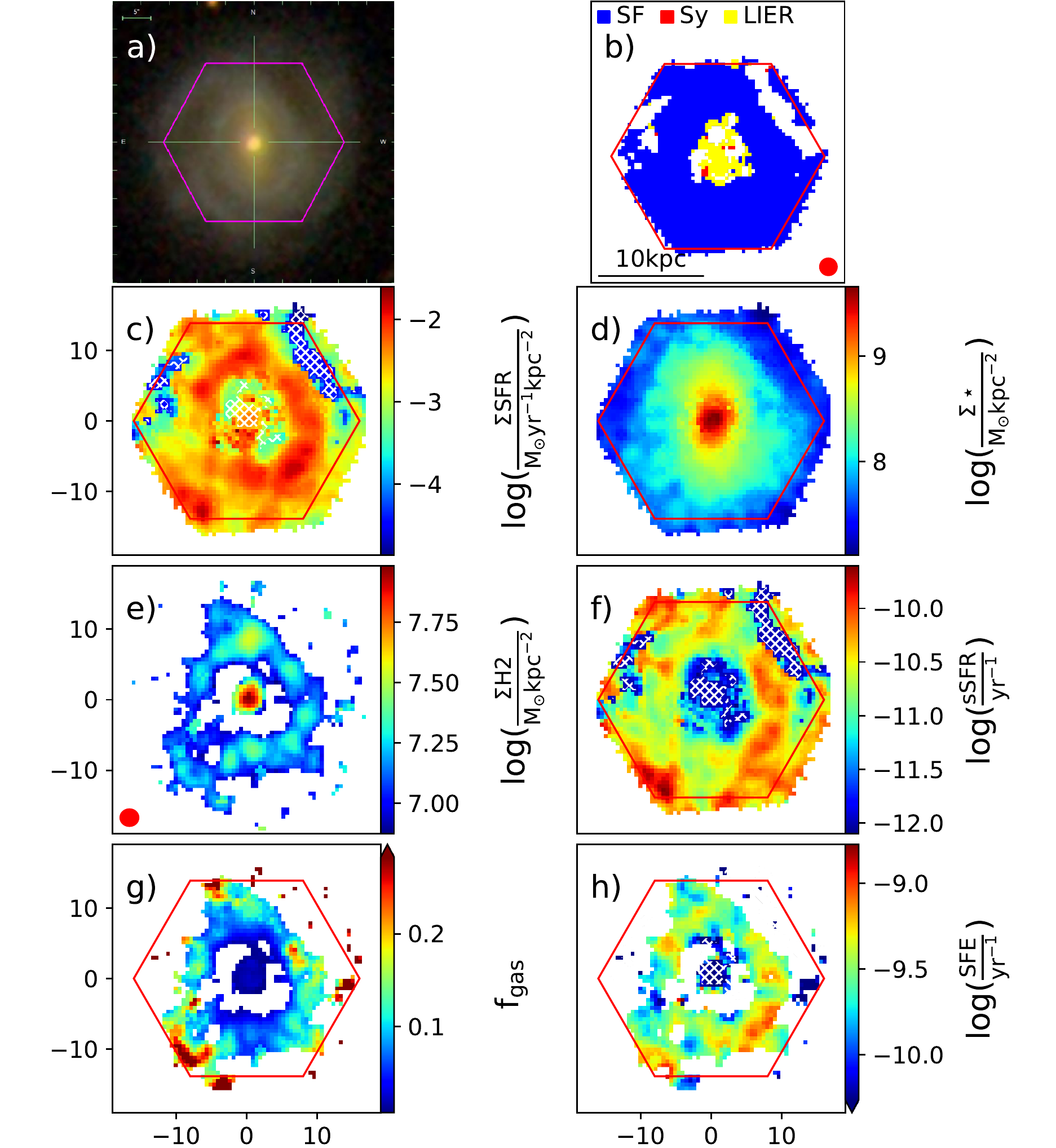}
    \caption{Resolved maps of galaxy 8550-12704. a) SDSS \textit{g,r,i} composite images. b) BPT classification, where blue, red and yellow correspond to star-forming, Seyfert and LIER, respectively, and the ellipse represents the MaNGA PSF. c) $\rm \Sigma_{SFR}$ estimated using the H$\alpha$ flux and full spectral fiting. d) $\rm \Sigma_{\star}$ from spectral fitting. e) $\rm \Sigma_{H_{2}}$ from NOEMA \textsuperscript{12}CO(1-0) observations, with the ellipse in the lower left corner representing the synthesised beam. f) $\rm {sSFR}$. g) $\rm {f_{gas}}$. h)  $\rm {SFE}$. The magenta hexagon in all panels represents the MaNGA FoV. Hatching in panels c, d and f  indicates regions with no evidence of recent star formation, either in emission lines or young SSP templates, where we use our $\rm {sSFR}$ detection limit to constrain $\rm \Sigma_{SFR}$.} 
        \label{fig:ExampleResolvedMaps}
\end{figure}

In Fig. \ref{fig:ExampleResolvedMaps} we show maps of $\rm {SFE}$ and $\rm {f_{gas}}$ for galaxy 8550-12704. We show maps obtained by deriving physical properties on spaxel-by-spaxel basis purely for display purposes. All the results presented later in this section are based on binning in radial annuli. Fig. \ref{fig:ExampleResolvedMaps} demonstrates the centrally-suppressed sSFR typical of GV galaxies, and a central decrease in both $\rm {f_{gas}}$ and $\rm {SFE}$. The other six galaxies show qualitatively similar trends (see section B of the online supplementary material).

\subsection{Radial Profiles}
\label{Radial Analysis}

Fig. \ref{fig:ExampleResolvedMaps} highlights the limitations and challenges of a fully resolved analysis: we obtain a biased view of the galaxy by restricting our analysis to pixels where one of the two key tracers (SFR or $\rm M_{H_{2}}$) is well-detected. For example, low-$\rm {f_{gas}}$ regions may be hidden in Fig. \ref{fig:ExampleResolvedMaps} because of molecular gas non-detections. We therefore derive radial profiles based on the annular-averaged spectra described in the previous section to get a comprehensive view of GV galaxies.

\begin{figure}
\center
    \includegraphics[width=\columnwidth]{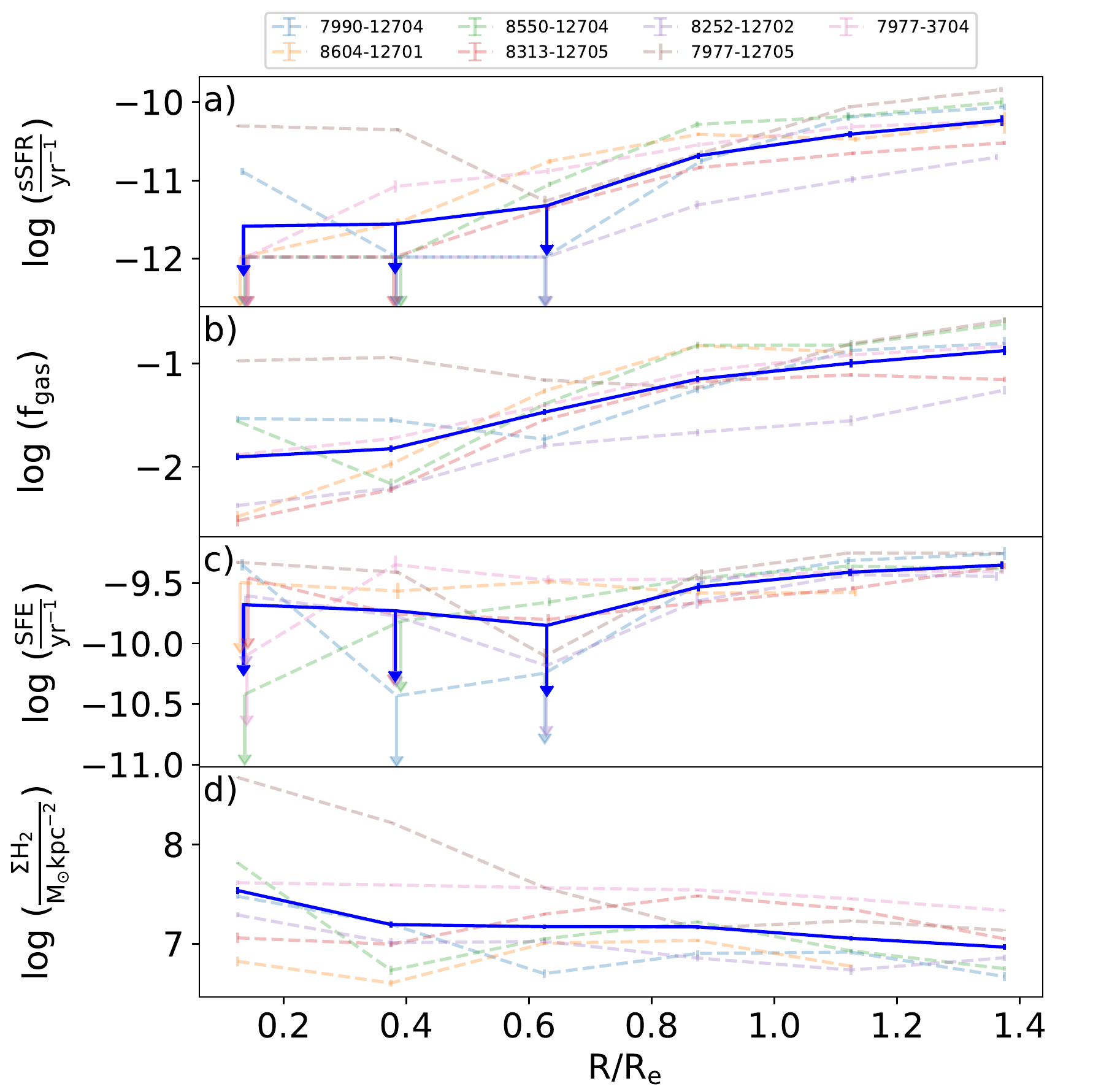}
    \caption{Radial profiles of $\rm sSFR$ (panel a), $\rm f_{gas}$ (panel b) and $\rm SFE$ (panel c) for each galaxy and the sample mean (solid blue profiles).  Many of the central annuli have no signs of recent star formation, so we use the  log($\rm sSFR/yr^{-1}) \sim -12 $  detection limit. The $\rm \Sigma_{H_{2}}$ radial profiles in panel d) show that central suppression of $\rm f_{gas}$ is not driven by a depleted gas reservoir.}
        \label{fig:AbsoluteRadialProfiles}
\end{figure}

Fig. \ref{fig:AbsoluteRadialProfiles} shows radial profiles of $\rm {sSFR}$, $\rm {f_{gas}}$ and $\rm {SFE}$ for the galaxies in our sample. The radial bins have widths of 0.25 $\rm R_e$, which corresponds approximately to the sizes of the MaNGA and NOEMA beams. 7977-3704 is an exception, where 0.25 $\rm R_e$ only corresponds to half the size of the NOEMA/MaNGA beam. The radial bins in this galaxy are therefore not independent, and the profiles are better viewed as moving averages.

$\rm {sSFR}$  profiles are consistent with previous GV studies  (e.g. \citealt{Spindler2018, Belfiore2018}): the sSFR shows a clear decrease moving from the outer to the inner regions, most of which is driven by the increase in $\rm \Sigma_{\star}$ in the central regions. We also find that the $\rm {SFE}$ shows a radial gradient, being lower at smaller galactocentric distances. This suggests that the decreasing $\rm {sSFR}$ is not only due to a stellar bulge. Rather, star formation is being suppressed. In agreement with \citet{Lin2017}, we observe reduced gas fractions in the central regions of GV galaxies with respect to their outskirts. The reduction is very significant (1 dex on average), except in 7977-12705. In this galaxy the central star formation is at least partially driven by a large gas reservoir. 

It is difficult to infer the full extent of any central suppression in $\rm {SFE}$ since our $\rm {sSFR}$ detection limit leads to $\rm \Sigma_{SFR}$ upper limits which are not strongly constraining in the high $\rm \Sigma_{\star}$ central regions. In fact, the central $\rm {SFE}$ upper limits are consistent with a relatively flat profile as well as a rapidly decreasing efficiency at small radii. Nonetheless, star formation tends to be less efficient at small galactocentric radii, and this effect compounds the reduction of $\rm {f_{gas}}$ to leave central regions quenched i.e. $\rm {sSFR}~\sim 10^{-12}~{yr}^{-1}$.

Assuming a metallicity-dependent $\rm \alpha_{CO}$ conversion factor in the presence of a metallicity gradient would reduce the measured central gas densities, thereby steepening radial profiles of $\rm {f_{gas}}$ whilst flattening those of $\rm {SFE}$. In the outer, star-forming regions, where the gas phase metallicity can be measured using standard diagnostics (here we use the O3N2 calibration from \citealt{Pettini2004}), the metallicity profiles are flat ($\rm 8.6<12+log(O/H)<8.8$) with $\rm 1\sigma~$ scatter smaller than 0.05 for all seven galaxies. Assuming the metallicity profiles remain flat in the LIER regions, where the metallicity cannot be directly measured, we expect $\rm \alpha_{CO}$ variations smaller than $\rm \sim 0.1~dex$ using the metallicity-dependent conversion factor adopted in \cite{Sun2020}.  This is insufficient to alter the trends shown in Figs. \ref{fig:AbsoluteRadialProfiles} and \ref{fig:OffsetRadialProfiles}.

\subsection{What is driving the reduced sSFR in GV galaxies?}
\label{key driver}

\begin{figure}
\center
    \includegraphics[width=\columnwidth]{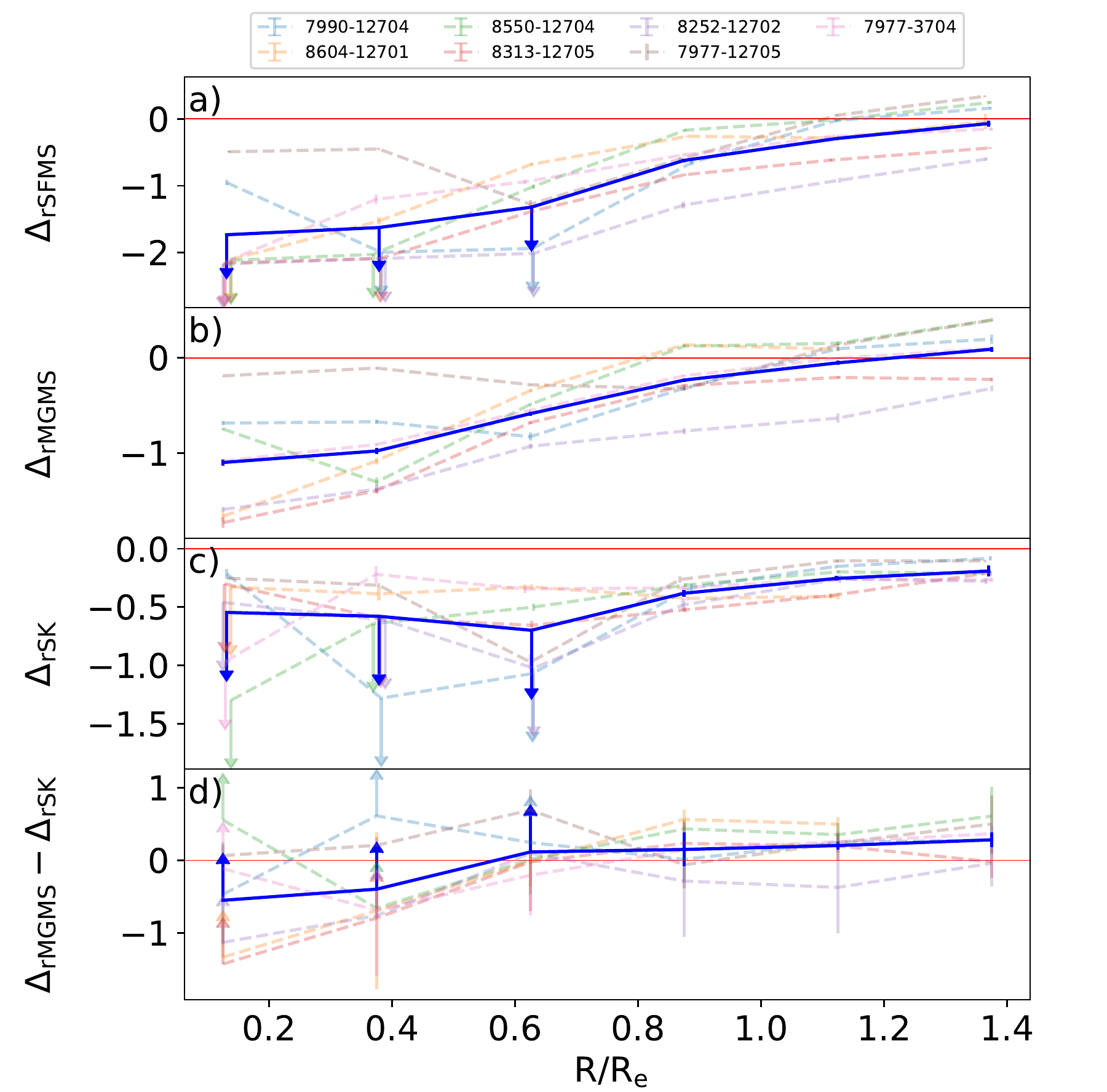}
    \caption{Radial profiles of offsets from the main sequence relationships presented in \citep{Lin2019}: Offsets from the rSFMS ($\rm \Delta_{rSFMS}$; panel a), rMGMS ($\rm \Delta_{rMGMS}$; panel b) and rSK ($\rm \Delta_{rSK}$; panel c). All offsets are calculated in logarithmic space and are therefore dimensionless. Panel d, $\rm \Delta_{rMGMS} - \Delta_{rSK}$, compares offsets from the rMGMS and rSK to rank the two drivers.}
        \label{fig:OffsetRadialProfiles}
\end{figure}

We have shown that  $\rm {f_{gas}}$ and $\rm {SFE}$ vary within GV galaxies and that both effects drive reductions in $\rm {sSFR}$. The key goal of this work is, however, to investigate the transition from the star-forming main sequence to the GV, and understand why GV galaxies form fewer stars than their star-forming counterparts. We therefore examine offsets from three relationships connecting $\rm \Sigma_{\star}$, $\rm \Sigma_{SFR}$  and $\rm \Sigma_{H_{2}}$ in MS galaxies on kpc scales, i.e. the rSFMS, rMGMS and rSK \citep{Lin2019} (Fig. \ref{fig:OffsetRadialProfiles}). We correct for the different IMF used in \citet{Lin2019} (Salpeter) when calculating the offsets.

GV galaxies lie below the rSFMS at all radii, but the difference is largest inside 0.5~$\rm R_{e}$, where the LIER regions form stars $\sim$100 times more slowly than star-forming regions. This is partially explained by offsets from the rMGMS: for the same $\rm \Sigma_{\star}$, the gas fraction is slightly reduced in the disc but $\sim$10 times lower in the inner bulge (relative to the gas fraction found in typical annuli in star-forming galaxies).  These offsets are at least partially driven by the growth of the central bulge and may not be associated with a change in the gas content of the disc. We also observe offsets from the rSK, and a mild radial trend. In particular, the efficiency of forming stars relative to normal star-forming galaxies also decreases with decreasing galactocentric radius. As with profiles of $\rm {SFE}$, much of the suppression lies below the detection limit. Nonetheless, we demonstrate that the efficiency of star formation at the centre of GV galaxies is generally three times lower than expected from the rSK, and some galaxies are up to 10 times less efficient. This confirms that the offset from the rSFMS is not only caused by the growth of the central bulge and that star-formation is suppressed at the centres of GV galaxies.

The bottom row in Fig. \ref{fig:OffsetRadialProfiles} compares offsets from the rMGMS and rSK relationships and enables a ranking of the two drivers: which is more significant for quenching star formation? Neither factor dominates beyond $\rm \sim$0.6~R$\rm_{e}$, and we conclude that changes in the gas reservoir and efficiency are equally responsible for reduced star formation in the disc. Offsets from the rMGMS appear to dominate in the central regions, but the full extents of the corresponding offsets from the rSK are unconstrained. We are therefore unable to rank the two drivers in these regions.

\begin{figure}
\center
    \includegraphics[width=\columnwidth]{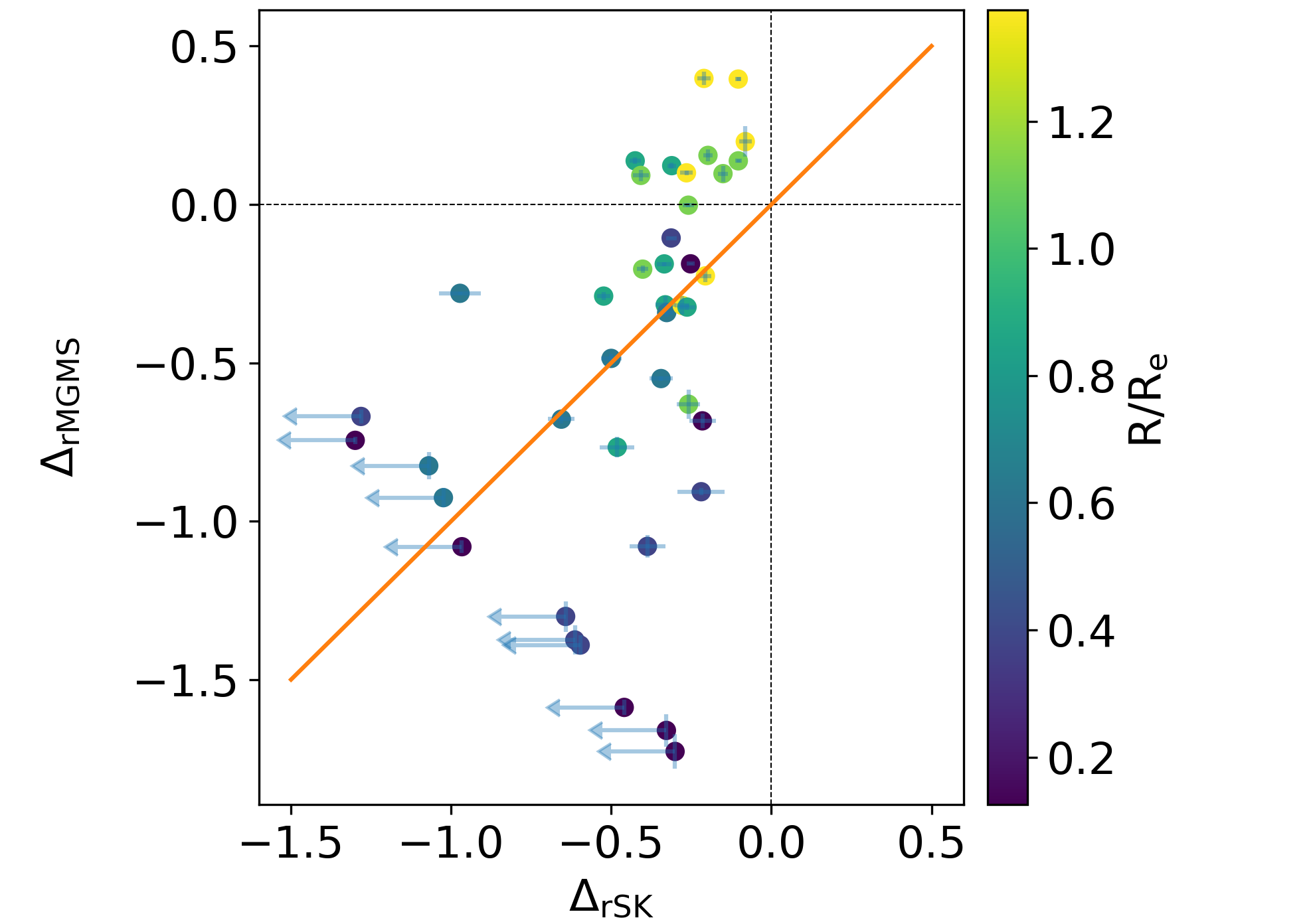}
    \caption{A comparison of offsets from the rMGMS and rSK for all annuli in the sample, colour-coded by galactocentric radius. The  $\rm \Delta_{rMGMS} = \Delta_{rSK}$ line is shown in orange.}
        \label{fig:DeltaMGMSVSDeltaSK}
\end{figure}

We review the offsets from the rMGMS and rSK for all annuli in Fig. \ref{fig:DeltaMGMSVSDeltaSK}. The majority of data points lie in the bottom left quadrant with suppressed gas reservoirs and star-forming efficiencies. Small galactocentric radii have significantly suppressed gas fractions and star formation efficiencies, but the upper limits on $\rm \Delta_{rSK}$ highlight our inability to identify the dominant mechanism. Whilst these data points generally lie below the 1:1 line where suppression of the gas reservoir dominates, our data is also consistent with the remarkable scenario in which GV galaxies are predominately quenched through reduced $\rm {SFE}$.

\section{Summary and Discussion}

We investigated the spatial distribution of molecular gas and star formation on kpc scales within GV galaxies. We find that both $\rm {f_{gas}}$ and $\rm {SFE}$ drive quenching. In particular, they are roughly equally responsible for quenching star formation in the outer disc. We are unable to determine the dominant mechanism in the central, strongly quenched regions, because of the difficulty to measure low levels of star formation, below roughly $\rm log( SFR/{yr}^{-1})\sim -12 $; but the data demonstrates that both $\rm {f_{gas}}$ and  $\rm {SFE}$ certainly contribute.

Our analysis is consistent with the results of global studies in which  $\rm SFE$ and $\rm f_{gas}$ both regulate $\rm sSFR$ \citep{Saintonge2011a, Saintonge2011b, Saintonge2017, Huang2014, Piotrowska2019, Zhang2019}.  Although  $\rm f_{gas}$  is the main driver of offsets from the global MS \citep{Saintonge2012}, the distribution of galaxies in the $\rm SFR-M_{\star}$  plane also depends on  variations in $\rm SFE$  \citep{Saintonge2016}. Our resolved analysis is consistent with $\rm {f_{gas}}$ driving quenching in the central regions of GV galaxies, but we have not ruled out  a scenario in which $\rm {SFE}$ is significantly suppressed and is the main cause of quenching.

A number of mechanisms may reduce the central $\rm {f_{gas}}$. The \textsc{simba} hydrodynamical simulation requires an AGN ejective mode to reproduce the central suppression in $\rm {sSFR}$ observed in GV galaxies, a success other simulations like \textsc{illustris} and \textsc{eagle} have not yet achieved \citep{Appleby2020}. However, quenching in \textsc{simba} is driven by $\rm {f_{gas}}$  in inner regions and  $\rm {SFE}$  in the outskirts. This is inconsistent with our findings. From an observational perspective, whilst it is natural to invoke large scale AGN-driven outflows to expel gas and reduce $\rm {f_{gas}}$ \citep{Maiolino2012}, molecular outflow velocities are generally found to be below the escape velocity, therefore raising doubts about their quenching ability \citep{Fluetsch2019}. Furthermore, much of the suppression in $\rm {f_{gas}}$ is driven by the large central bulge rather than reduced gas content (Fig. \ref{fig:AbsoluteRadialProfiles} panel d). Radial profiles of $\rm \Sigma_{H_{2}}$ tend to increase slightly with decreasing galactocentric radius. This may point towards preventive, rather than ejective feedback. Star-forming galaxies, with their centrally elevated $\rm \Sigma_{SFR}$, will subsequently build up their central $\rm \Sigma_{\star}$, decreasing $\rm {f_{gas}}$. Thus centrally suppressed $\rm {f_{gas}}$ may simply be a consequence of star formation in a galaxy starved of its gas supply.

AGN may also suppress $\rm {SFE}$ by injecting thermal energy directly into the ISM and supporting molecular clouds against gravitational collapse. Magnetic fields and turbulence may provide alternative sources of pressure support \citep{Federrath2012}. Finally, the galaxies in our sample have prominent bulges, which may support the disc against gravitational instabilities and suppress $\rm {SFE}$ \citep{Martig2009}.

\cite{James2009} discuss the possibility of bars sweeping out `star formation deserts', often accompanied by excess star formation at the centre of the bar. 7977-12705 and 7990-12704 show the clearest evidence of bars in our sample, and both have the largest central $\rm {sSFR}$ (Fig. \ref{fig:AbsoluteRadialProfiles}). Whilst these central regions have $\rm {SFE}$ consistent with the other five galaxies, they have increased $\rm {f_{gas}}$, supporting a scenario in which bars encourage the inflow of gas towards a galaxy's centre \citep{Regan2004}.

It is tempting to assess the two drivers, $\rm {f_{gas}}$ and $\rm {SFE}$, by comparing their correlations with $\rm {sSFR}$ \citep{Lin2017, Ellison2019b}. This approach runs into two potential issues. Firstly, it relies on constraining all parameters ($\rm {sSFR}$, $\rm {f_{gas}}$ and $\rm {SFE}$) throughout the galaxy. \citet{Lin2017}, on the other hand, only consider star-forming regions that have emission line and \textsuperscript{12}CO(1-0) fluxes exceeding the detection limits. This biases the results towards less quenched regions. Though we have improved the analysis by constraining all radii, some measurements of $\rm {SFE}$ are upper limits that cannot trivially be included in a correlation analysis. Secondly, the three derived parameters actually rely on only two independent measurements: $\rm {sSFR}$ and $\rm {SFE}$ both include a $\rm \Sigma_{SFR}$ term, and $\rm {sSFR}$ and $\rm {f_{gas}}$ both include a $\rm \Sigma_{\star}$ term. Strong correlations are therefore to be expected. In fact, the strength of the correlation increases as the confounding measurements become more noisy. Correlation analyses should therefore be treated with caution. 

In the near future the ALMaQUEST sample will be further expanded, allowing the study of secondary correlations (e.g. the role of stellar mass). HCN observations will also be forthcoming, aimed at directly investigating the true site of star formation: dense molecular gas \citep{Gao2004}.

\section*{Acknowledgements}
This work is based on observations carried out with the IRAM Plateau de Bure Interferometer. IRAM is supported by INSU/CNRS (France), MPG (Germany) and IGN (Spain). 
This paper makes use of the following ALMA data:ADS/JAO.ALMA\#2015.1.01225.S. ALMA is a partnership of ESO(representing its member states), NSF (USA) and NINS (Japan), together with NRC (Canada), MOST and ASIAA (Taiwan), and KASI (Republic of Korea), in cooperation with the Republic of Chile. The Joint ALMA Observatory is operated by ESO, AUI/NRAO and NAOJ.
RM acknowledges ERC Advanced Grant 695671 "QUENCH". SB and RM acknowledge
support by the Science and Technology Facilities Council (STFC). SC acknowledges support from the ERC Advanced Grant INTERSTELLAR H2020/740120. Funding for the Sloan Digital Sky Survey IV has been provided by the Alfred P. Sloan Foundation, the U.S. Department of Energy Office of Science, and the Participating Institutions. SDSS-IV acknowledges support and resources from the Center for High- Performance Computing at the University of Utah. The SDSS web site is http://www.sdss.org. SDSS-IV is managed by the Astrophysical Research Consortium for the Participating Institutions.




\bibliographystyle{mnras}
\bibliography{BibFiles/GreenValleyPaper}




\appendix
\section{Global Galaxy Properties}
\begin{table*}
 \renewcommand*{\arraystretch}{0.8}
\caption{Properties of the galaxies observed with NOEMA (top five rows) and ALMA (bottom two rows). Stellar masses, SFR and molecular gas masses are calculated within the MaNGA FoV.}
\begin{tabular}{c c c c c c c c c c c} 
 \hline
Plate-IFU & R.A. & DEC & Redshift & $\rm R_{e}$ & log($\rm \frac{M_{\star}}{M_{\odot}}$) & log($\rm \frac{SFR}{M_{\odot}yr^{-1}}$)  & log($\rm\frac{M_{H2}}{M_{\odot}}$) &  Beam Size  & 1$\rm \sigma$ noise \\
&    &  &  & (kpc)   &  &  & &    (arcsec $\times$ arcsec) &  (mJy~beam$^{-1}$~channel$^{-1}$)\\
\hline
7990-12704  & 17:29:56.64 & 58:23:50.68 & 0.02682 & 5.36 & 10.29  & -0.48 & 8.98 &  (2.46 $\times$ 2.04) & 0.80 \\
8252-12702 & 09:42:07.39 & 48:09:17.54 & 0.03367 & 5.71 & 10.90 & -0.51 & 9.08 &   (2.68 $\times$ 2.17) & 0.73 \\
8604-12701 & 16:23:32.74 & 39:07:15.91 & 0.03504 & 8.70 & 10.85  & -0.11 & 9.42 &   (3.61 $\times$ 2.80) & 1.10 \\
8550-12704  & 16:28:14.01 &  40:18:49.84 & 0.03311 & 6.06 & 10.66 & -0.15 & 9.38 &  (2.54 $\times$ 2.36) & 0.78 \\
8313-12705  & 16:10:43.81 & 41:08:54.82 & 0.03154 & 6.05 &10.93 & -0.13 & 9.44 &  (2.51 $\times$ 2.35) &   0.81 \\
7977-12705  &  22:11:34.27  &   11:47:45.24   &  0.02695 & 4.77 & 10.68 & 0.31 & 9.70 & (2.80 $\times$ 2.36)  & 0.64   \\
7977-3704  &  22:11:11.70  &  11:48:02.64  & 0.02702 & 2.16 & 10.21 & -0.56 & 8.88 & (2.64 $\times$ 2.34) &   0.76  \\
\hline
 \end{tabular}
\label{table:GalaxyProerties}
\end{table*}

\section{Spatially Resolved Maps}
\begin{figure}
\center
    \includegraphics[width=0.5\textwidth]{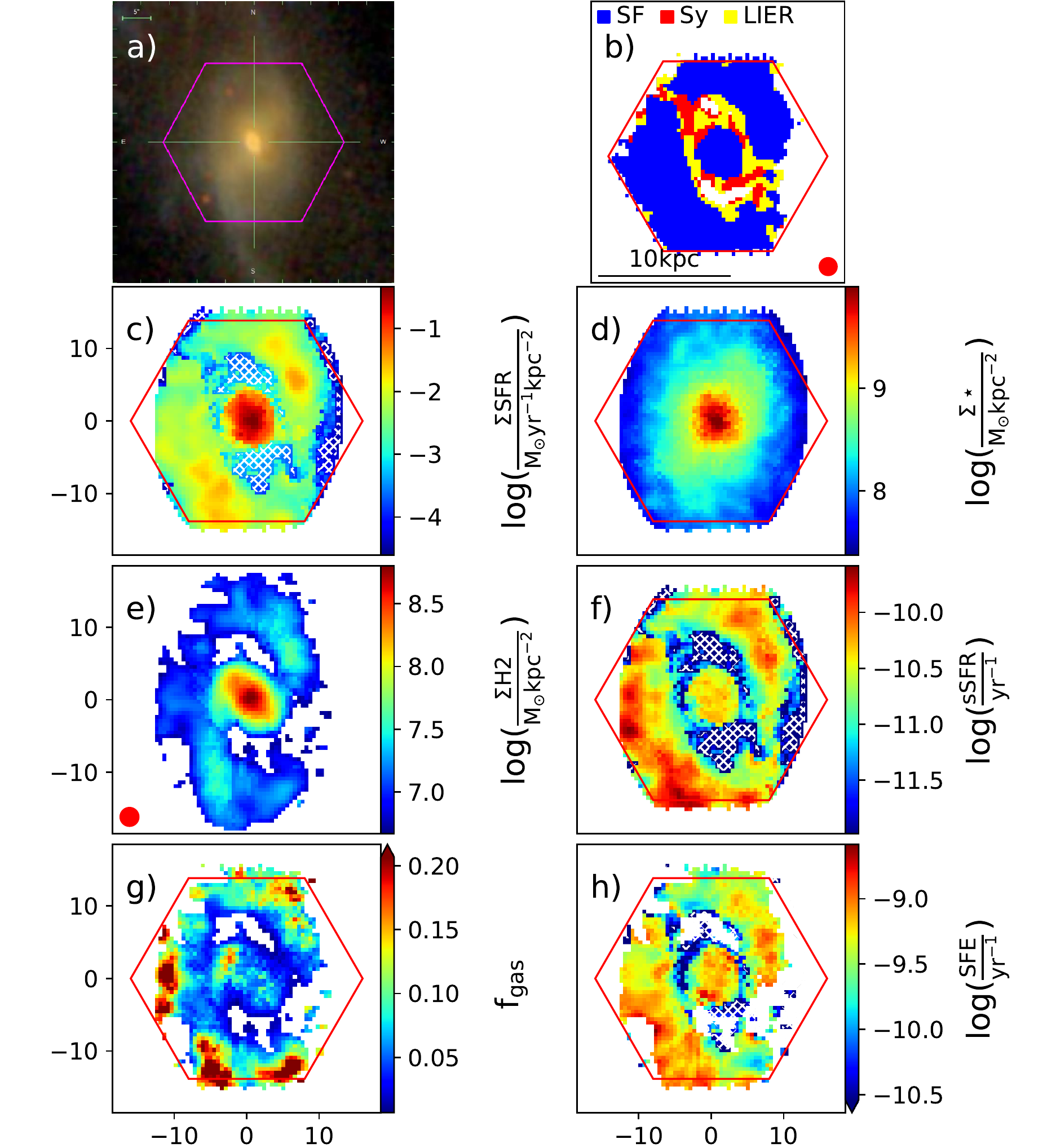}
    \caption{7977-12705} 
        \label{fig:7977-12705Maps}
\end{figure}

\begin{figure}
\center
    \includegraphics[width=0.5\textwidth]{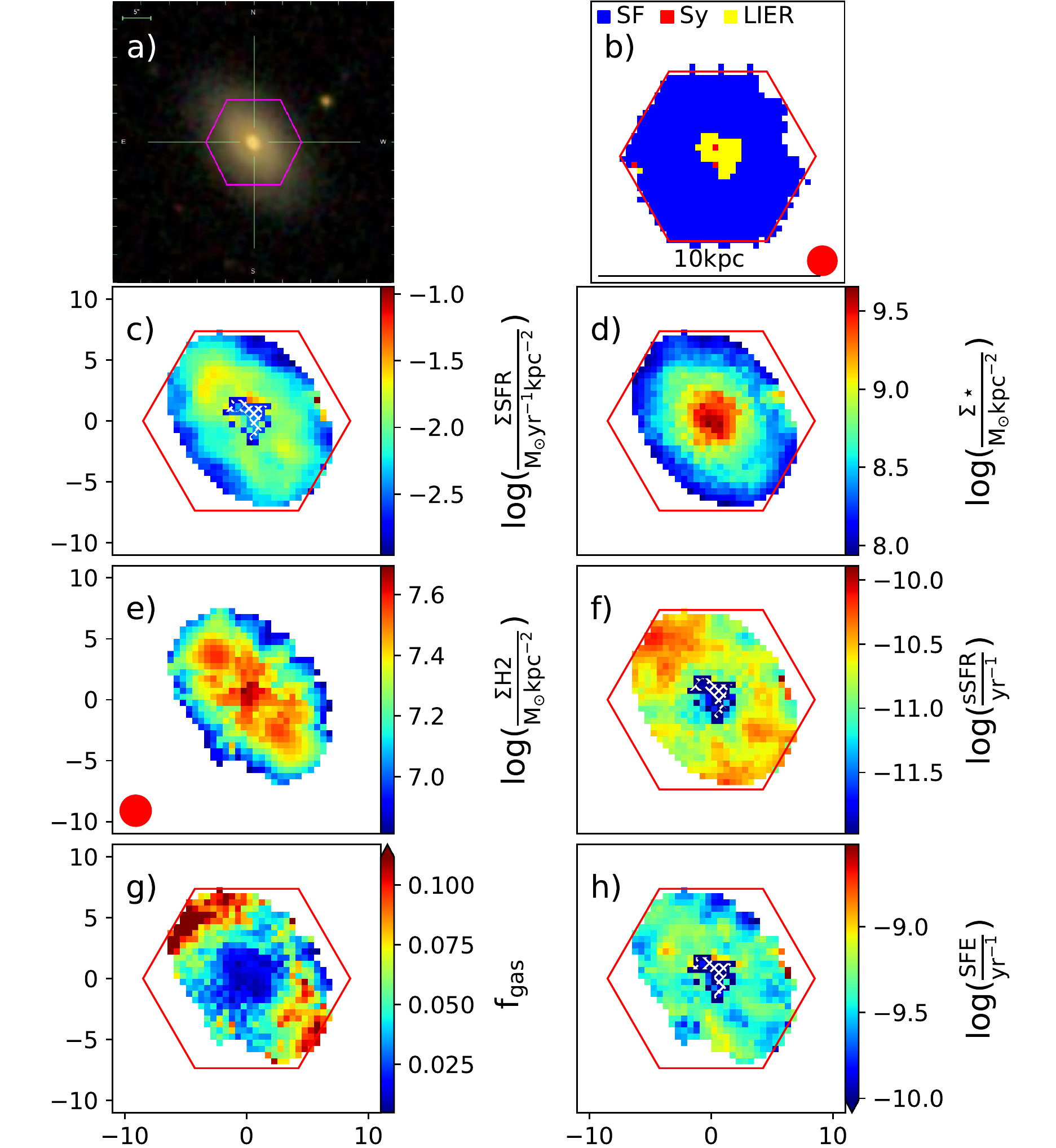}
    \caption{7977-3704} 
        \label{fig:7977-3704Maps}
\end{figure}

\begin{figure}
\center
    \includegraphics[width=0.5\textwidth]{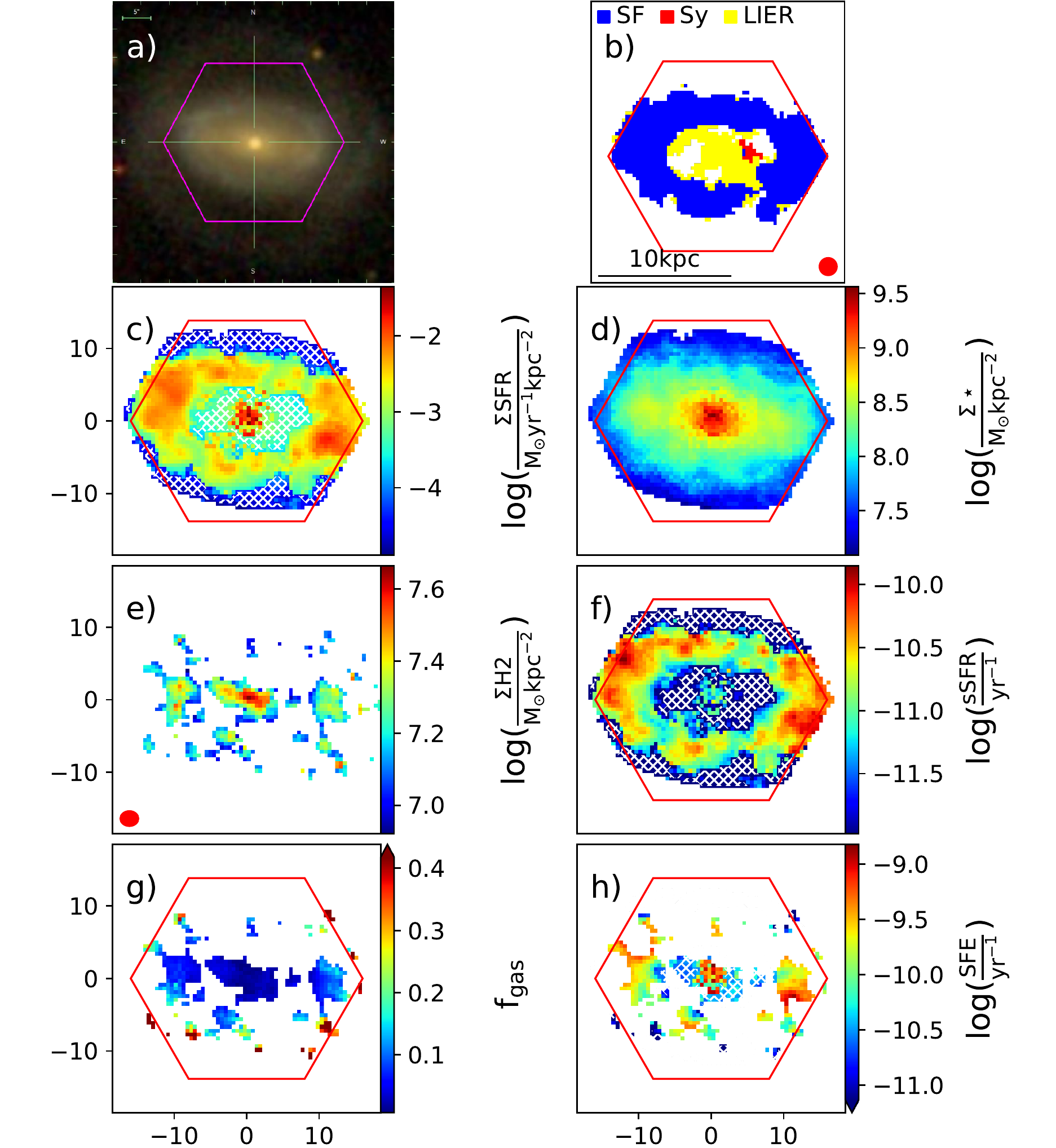}
    \caption{7990-12704} 
        \label{fig:7990-12704Maps}
\end{figure}

\begin{figure}
\center
    \includegraphics[width=0.5\textwidth]{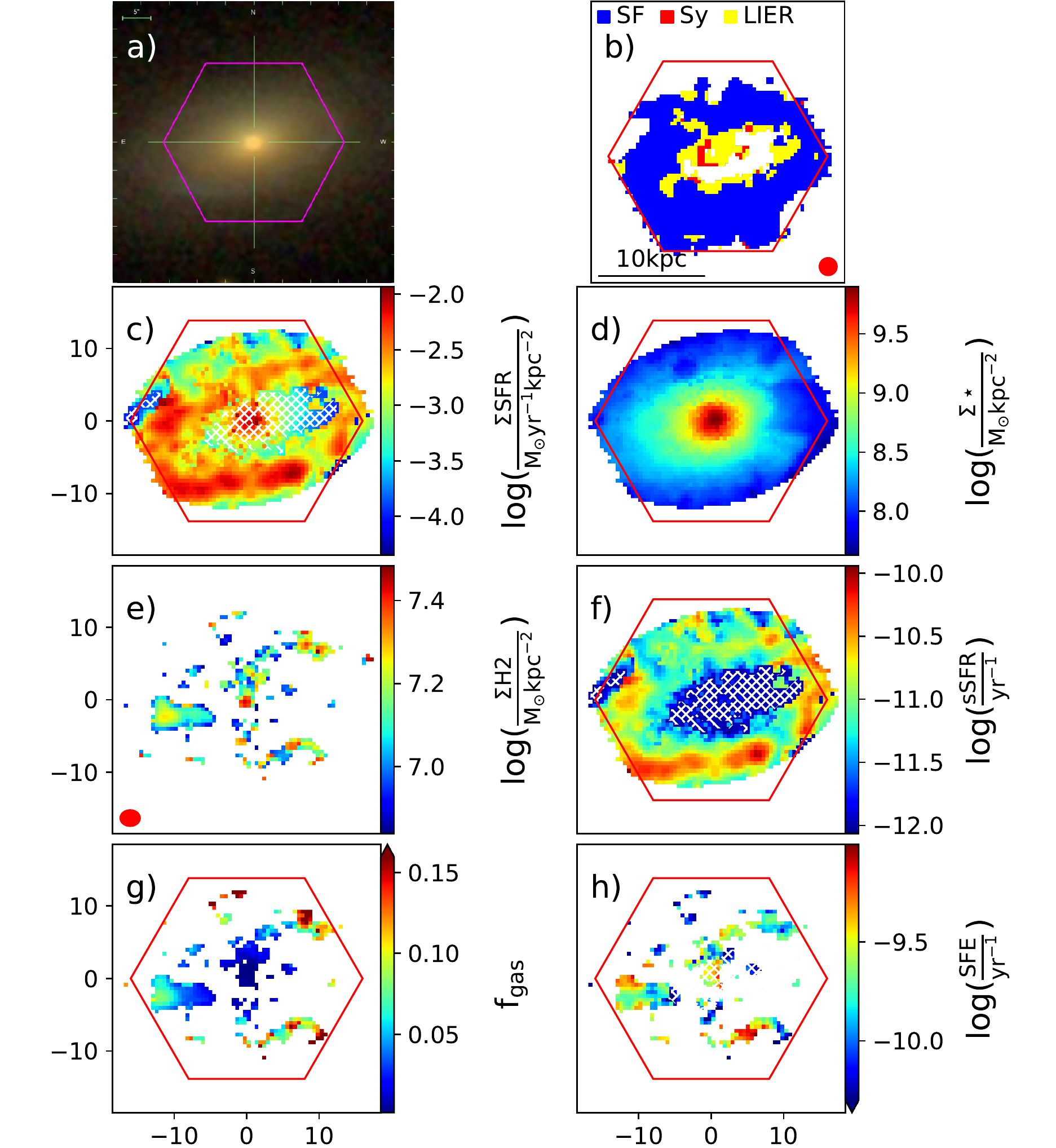}
    \caption{8252-12702} 
        \label{fig:8252-12702Maps}
\end{figure}

\begin{figure}
\center
    \includegraphics[width=0.5\textwidth]{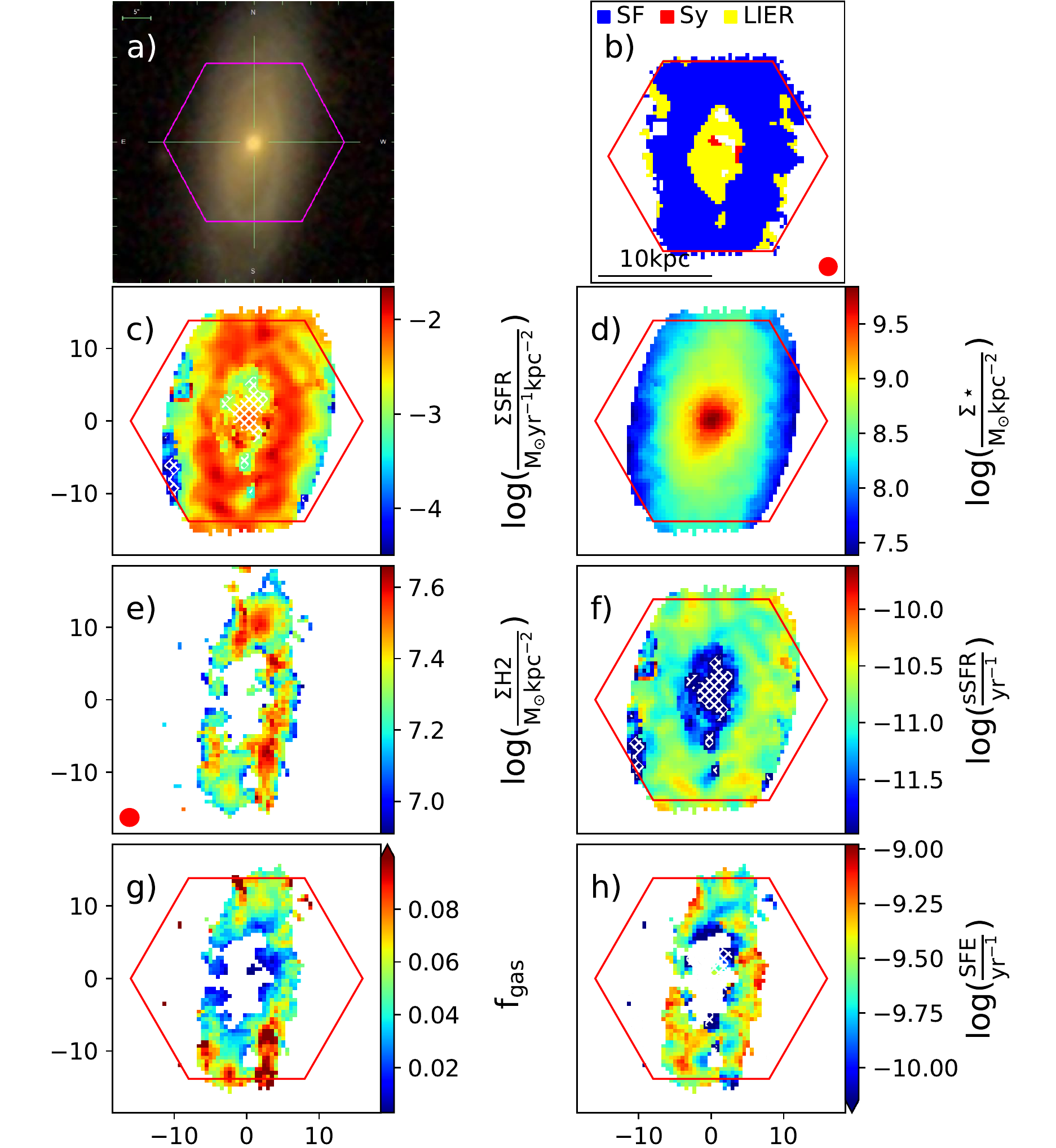}
    \caption{8313-12705} 
        \label{fig:8313-12705Maps}
\end{figure}

\begin{figure}
\center
    \includegraphics[width=0.5\textwidth]{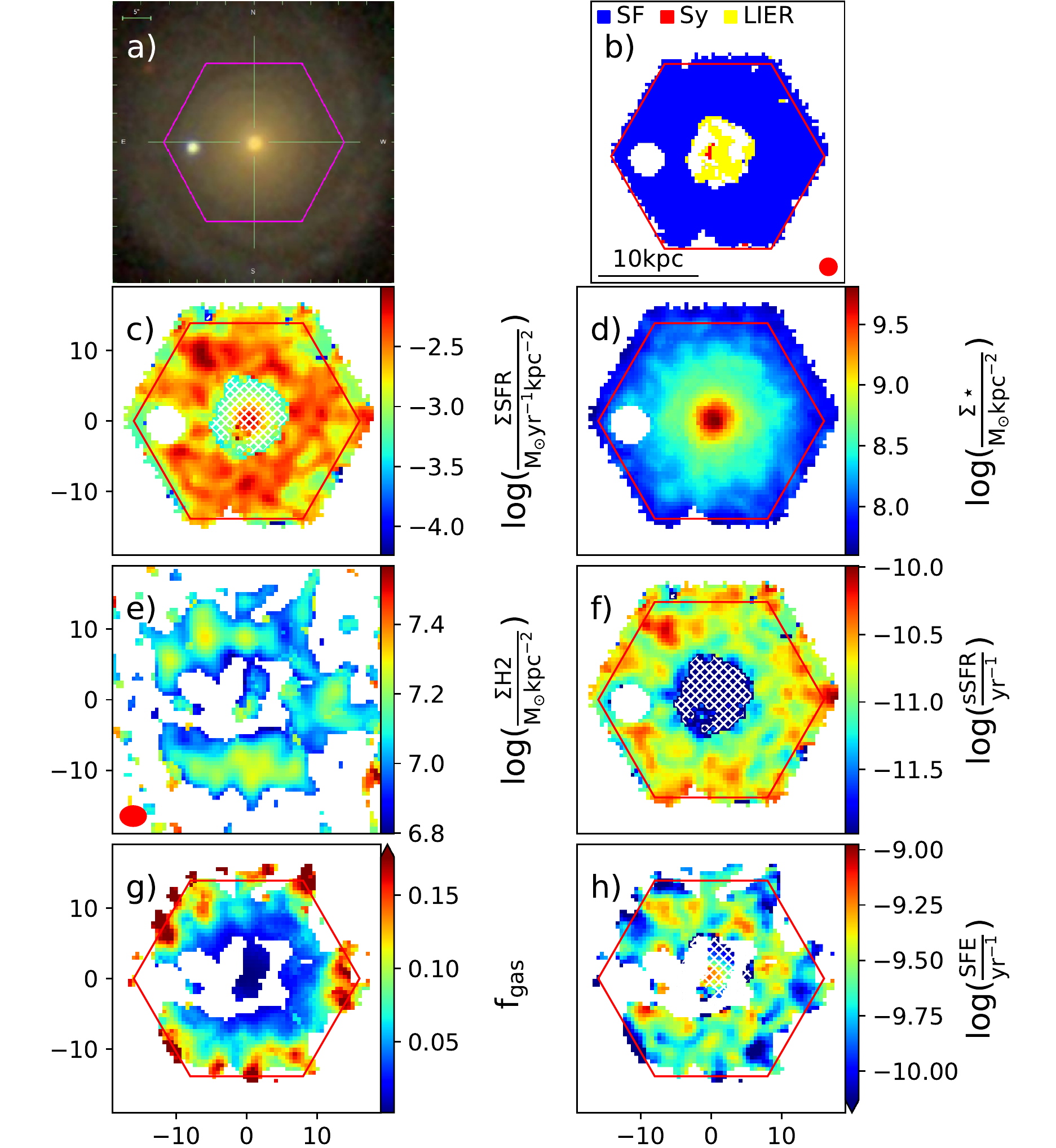}
    \caption{8604-12701} 
        \label{fig:8604-12701Maps}
\end{figure}

In Figs. \ref{fig:7977-12705Maps}, \ref{fig:7977-3704Maps}, \ref{fig:7990-12704Maps}, \ref{fig:8252-12702Maps}, \ref{fig:8313-12705Maps} and \ref{fig:8604-12701Maps} we show the following maps: a) SDSS \textit{g,r,i} composite images. b) BPT classification, where blue, red and yellow correspond to star forming, Seyfert and LIER, respectively, and the ellipse represents the MaNGA PSF. c) $\rm \Sigma_{SFR}$ estimated using the H$\alpha$ flux and full spectral fiting. d) $\rm \Sigma_{\star}$ from spectral fitting. e) $\rm \Sigma_{H_{2}}$ from NOEMA \textsuperscript{12}CO(1-0) observations, with the ellipse in the lower left corner representing the synthesised beam. f) $\rm {sSFR}$. g) $\rm {f_{gas}}$. h)  $\rm {SFE}$. The magenta hexagon in all panels represents the MaNGA FoV. Hatching in panels c, d and f  indicates regions with no evidence of recent star formation, either in emission lines or young SSP templates. We use our $\rm {sSFR}$ detection limit to constrain $\rm \Sigma_{SFR}$ in these regions.


\section{Radially Binned \textsuperscript{12}CO(1-0) Spectra}
\begin{landscape}
\begin{figure}
\center
    \includegraphics[width=\columnwidth]{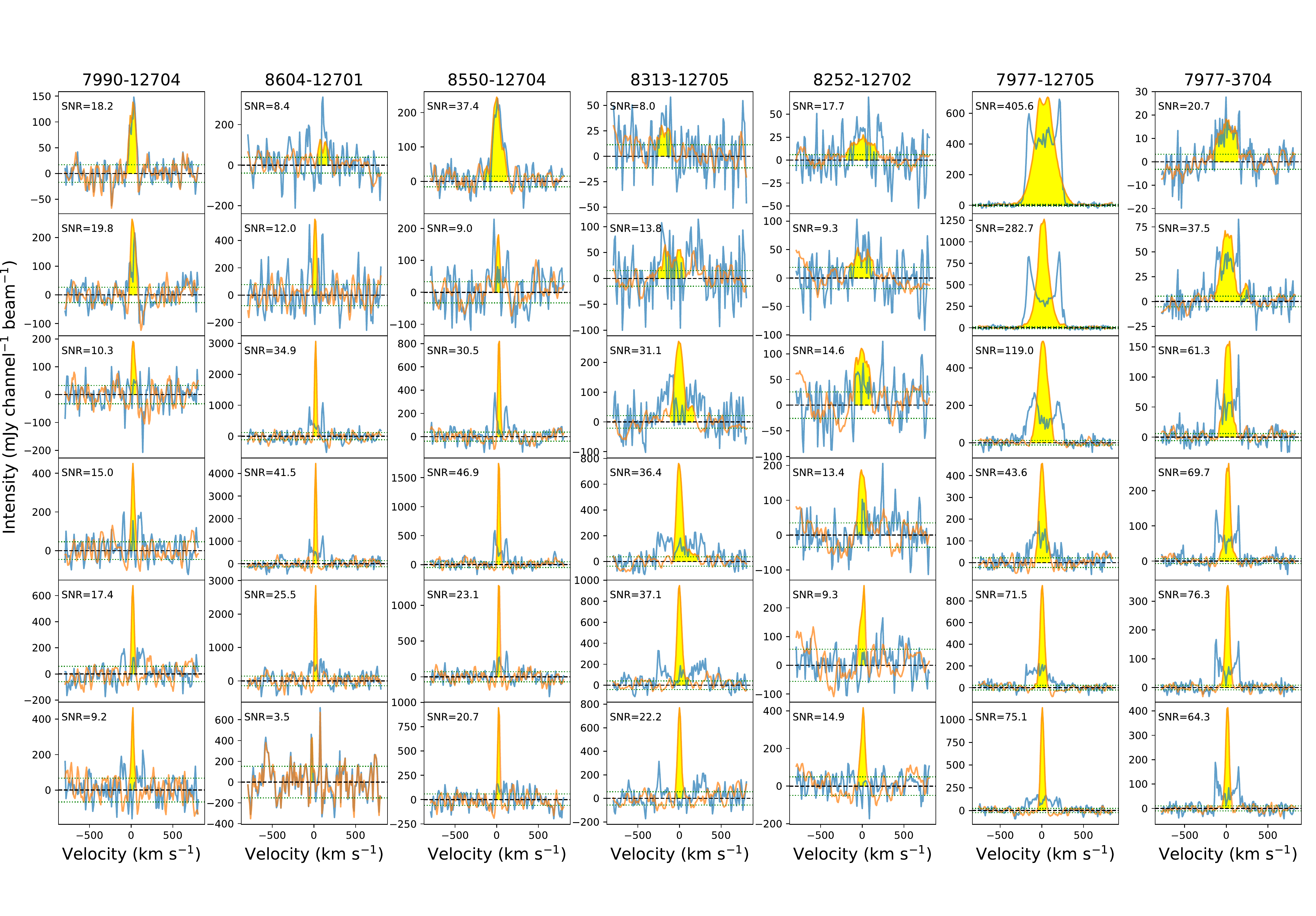}
    \caption{Stacked \textsuperscript{12}CO(1-0) in radial bins of width  0.25~$\rm R_{e}$, with the radial bins $\rm [0, 0.25]~R_{e}$ shown in the top row and $\rm [1.25, 1.5]~R_{e}$ in the bottom row. The blue spectra are achieved by coadding the spectra of the spaxels within each radial bin, whilst the orange spectra are found by recentering the spectra within each spaxel (using the $\rm H\alpha$ velocity field from the \textsc{dap}) before coadding. The shaded yellow areas represent the regions of the orange spectra used for estimating the \textsuperscript{12}CO(1-0) line flux. The dotted green lines mark the $\rm 1\sigma$ scatter in the line-free channels. This is the noise level used to calculate the SNR.} 
        \label{fig:BinnedSpectra}
\end{figure} 
\end{landscape}

Fig. \ref{fig:BinnedSpectra} shows stacked \textsuperscript{12}CO(1-0) spectra within six $\rm 0.25~R_{e}$ radial bins for each of the seven green valley galaxies. Recentering in each spaxel using the $\rm H\alpha$ velocity field from the \textsc{dap} increases the SNR of the \textsuperscript{12}CO(1-0) flux within each channel. This effect of removing the double-horn profile is well demonstrated in galaxy 7977-12705. In each panel we show the SNR of total \textsuperscript{12}CO(1-0) line flux. The 8604-12701 $\rm [1.25, 1.50]~R_{e}$ bin is the only one with SNR below five and is removed from the analysis. 

\label{lastpage}
\end{document}